# Assessing r²SCAN meta-GGA functional for structural parameters, cohesive energy, mechanical modulus and thermophysical properties of 3*d*, 4*d* and 5*d* transition metals

Haoliang Liu[a], Xue Bai[b], Jingliang Ning[c], Yuxuan Hou[a], Zifeng Song[a], Akilan Ramasamy[c], Ruiqi Zhang[c], Yefei Li[b*], Jianwei Sun[c*], Bing Xiao[a*]

[a] *State Key Laboratory of Electric Insulation and Power Equipment and School of Electrical Engineering, Xi'an Jiaotong University, Xi'an 710049, China*
[b] *State Key Laboratory for Mechanical Behavior of Materials, Xi'an Jiaotong University, Xi'an 710049, China*
[c] *Department of Physics and Engineering Physics, Tulane University, New Orleans, , United States*

* Correspondence authors: Yefei Li, E-mail: liyefei@xjtu.edu.cn; Jianwei Sun, E-mail: jsun@tulane.edu; Bing Xiao, E-mail: bingxiao84@xjtu.edu.cn

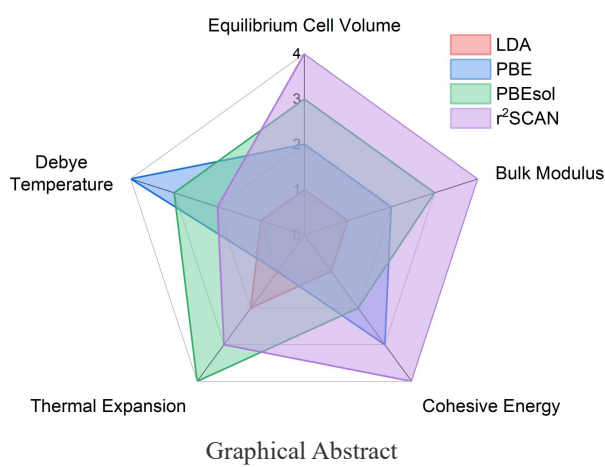

Graphical Abstract

**Abstract**

    The recent development of the accurate and efficient semilocal density functionals on the third rung of Jacob's ladder of density functional theory such as the revised regularized strongly constrained and appropriately normed (r²SCAN) density functional could enable the rapid and highly reliable prediction of the elasticity and temperature dependence of thermophysical parameters of refractory elements and their intermetallic compounds using quasi-harmonic approximation (QHA). Here, we present a comparative evaluation of the equilibrium cell volumes, cohesive energy, mechanical moduli, and thermophysical properties (Debye temperature and thermal expansion coefficient) for 22 transition metals using semilocal density functionals, including local density approximation (LDA), the Perdew-Burke-Ernzerhof (PBE) and PBEsol generalized gradient approximations (GGA), and the r²SCAN meta-GGA. PBEsol and r²SCAN deliver the same level of

accuracies for structural, mechanical and thermophysical properties. Otherwise, PBE and r$^2$SCAN perform better than LDA and PBEsol for calculating cohesive energies of transition metals. Among the tested density functionals, r$^2$SCAN provides an overall well-balanced performance for reliably computing the cell volumes, cohesive energies, mechanical properties, and thermophysical properties of various 3*d*, 4*d*, and 5*d* transition metals using QHA. Therefore, we recommend that r$^2$SCAN could be employed as a workhorse method to evaluate the thermophysical properties of transition metal compounds and alloys in the high throughput workflows.

1. Introduction

Refractory transition metals and their alloys or intermetallic compounds generally show excellent mechanical and thermophysical properties such as the high-melting points, high hardness, large mechanical moduli, high tensile or shearing strength, and superior wearing resistance. Therefore, they are the key ingredients for fabricating various super-alloys which are widely used to manufacture the automobile, aerospace, and industrial gas turbine engine components. A typical example of those alloys is the Ni-based alloys where the Cr, Ti, Al, Co, Fe, Nb, Mo, W, and Ta are usually added into the Ni base materials to enhance the mechanical strength and creep resistance at the operating temperature above 1000 $^o$C in gas turbines[1,2]. In recent years, a large number of high entropy alloys consisting of multiple refractory elements have also been widely investigated for their potential use on the gas path and aerospace engines[3]. Meanwhile, many refractory transition metals are also the vital ingredients in various advanced single atom catalysts for energy conversion or storage materials[4].

Nowadays, the high throughput first principles calculations are routinely employed in the rational designing of novel super-alloys and high entropy alloys with high efficiency and low carbon footprint, compared to the traditional high energy input experimental methodology. For super-alloys or high entropy alloys, their mechanical and thermophysical properties such as mechanical moduli, enthalpy of formation, heat capacity and volumetric thermal expansion coefficients play the vital role for the relevant applications at high operating temperature. Using the density functional theory (DFT) to confidently predict those aforementioned properties for refractory elements and their intermetallic compounds is still a challenging task at the moment, because the lacking of an universal and highly transferrable semilocal exchange-correlation functional to capture various types of electron-electron interactions in transition metals spanning from the early transition elements with the highly localized $3d$ orbitals to the late $4d$ and $5d$ elements with the largely delocalized d states. The deficiency of a semilocal exchange-correlation functional in describing the itinerant magnetism of transition metals with different degrees of localization for $d$ orbitals could also have great impact on the reliability of predicting the elasticity and thermophysical properties for the magnetic refractory elements and their intermetallic compounds[5,6].

Following the recommendation of Perdew and coworkers[7], the exchange-correlation density functionals using different ingredients for the construction are represented by the five rungs on the Jacobs' ladder. The first three rungs are all referred to the semilocal density functionals including local density approximation (LDA), generalized gradient approximation (GGA) and meta-generalized gradient approximation (meta-GGA). The fourth and fifth rungs include nonlocal ingredients and are typically computationally more expensive. Although, the overall accuracy of an exchange-correlation functional is improved by climbing up the Jacob's ladder of DFT, the very high computational costs and slow numerical convergence issue of the fourth and fifth rungs greatly limit their wide applications in either the high throughput calculations or for large complex crystal structures. For high throughout first principles calculations, the semi-local density functionals remain as the optimal methods in the near future. However, the most widely used semi-local functionals such as LDA and the Perdew-Burke-Ernzerhof (PBE) GGA do not provide the universal platform for describing the structural, mechanical and thermophysical properties with the same level of accuracy. It is now well-known that LDA usually predicts too small lattice constants for crystal structures and thus the elastic constants and mechanical moduli are usually overestimated[8,9]. On the other hand, PBE may overestimate the lattice parameters and thermal expansion coefficients, resulting in the underestimation of the mechanical moduli and elastic constants[10-12]. A fundamental dilemma of LDA and GGA functionals is that the structural properties (lattice constants, elastic constants and mechanical moduli) and energetic parameters (formation enthalpy and cohesive energy) cannot be accurately predicted simultaneously[13]. A very recent breakthrough is the development of the strongly constrained and appropriately normed (SCAN) meta-GGA functional[14]. Previous assessment clearly showed that this new meta-GGA level exchange-correlation density functional was able to describe a wide range of molecules and solids with great diversity in bond types, ranging from strong chemical bonds to hydrogen bonds, and to much weaker van der Waals interactions[15-20].

The reliability and numerical efficiency of SCAN meta-GGA functional have been assessed for a large number of transition metals and their alloys in some recent publications[19,21,22]. Specifically, Eric et al. calculated structural parameters and formation enthalpies of 945 compounds using the SCAN and PBE functionals, including many intermetallic alloys containing refractory transition metals[19]. All the above studies suggested that the SCAN functional could deliver the promising accuracy and reliability for calculating equilibrium lattice constants, elasticity and formation enthalpy for transition metals and their alloys. However, in the cases of the high degree of electron localization and magnetism, the good performance of SCAN is somehow compromised. For example, Nepal et al., employed PBE-GGA, SCAN-meta-GGA and RPA to calculate the lattice parameters and the mechanical moduli of Au-Cu solid solution alloys[6]. It was concluded that both SCAN and PBE are failed to provide the accurate bulk moduli for Au-Cu alloys, and only RPA offers the satisfactory results in the whole composition range in Au-Cu binary system. Another observed deficiency in SCAN functional is seen in the itinerant electron magnets[23-25]. Particularly, the structural parameters, local magnetic moments and ground state magnetic orderings of Fe, Co and Ni have been studied using SCAN functionals by Ekholm et al. in Ref [23] and Fu et al. in Ref [25]. Those investigations clearly concluded that the SCAN functional predicts too large local magnetic moments, and overstabilizes the ferromagnetic state.

Meanwhile, Bartok and Yates constructed the regularized SCAN (rSCAN) functional to directly address the numerically instability presented in the SCAN functional when the isoorbital indicator approaches to 1 for the slow varying density limit [26]. However, merely prioritizing numerical efficiency in rSCAN could result in the violation of many mathematical and physical construction principles which are respected by the original SCAN functional. To keep the numerical advances as well as address deficiencies in the construction principles of rSCAN, the r$^2$SCAN meta-GGA functional was proposed[27]. The primary tests on the structural and energetic properties of molecules and solids raised the high expectation that the r$^2$SCAN functional could provide not only state-of-the-art semilocal density functional in the high throughput first principles calculations, but also more importantly being considered as an universal platform to concomitantly predict structural, mechanical and thermophysical properties accurately for a wide range of alloys and intermetallic compounds.

Therefore, as strongly motivated by arguments above, we systematically calculated the equilibrium cell volumes, cohesive energy, bulk modulus, Debye temperature and volumetric thermal expansion coefficients for 22 pure refractory elements within 3$d$, 4$d$ and 5$d$ series using LDA, PBE, PBEsol, and r$^2$SCAN exchange-correlation functionals. Our calculations confirmed that r$^2$SCAN performs the best for cell volumes, cohesive energy, and mechanical properties among all tested semi-local density functionals (LDA, PBE, PBEsol and r$^2$SCAN). PBEsol (PBE) is slightly better than r$^2$SCAN for predicting the thermal expansion coefficient (Debye temperature). Therefore, r$^2$SCAN functional can be reliably used as the work horse functional in studying the structural, mechanical and thermophysical properties of various transition metals. It is expected that the promising performances of r$^2$SCAN could greatly benefit the theoretical investigation of intermetallic compounds and high entropy alloys in the relevant large-scale computational study.

**2. Computational methods and details**

All first principles calculations were performed using the plane wave basis under the three-dimensional periodic boundary conditions (PBCs) in a locally modified version of Vienna ab-initio software package (VASP 5.4.4) [28]. Both SCAN and r$^2$SCAN functionals were implemented in this version of VASP code. The projector augmented wave method was adopted to describe the interactions between ionic core and valence electrons [29]. The PBE-type PAW potentials for the relevant elements were employed in the current work (See Table SI for details). The kinetic energy cutoff value was set to 500 eV for the plane wave expansion in the reciprocal space. The Monkhorst-Pack scheme [30] was used for the numerical integrations in the first

Brillouin zone with a grid of 0.1 Å$^{-1}$ for all crystal structures. Within the current computational parameters, the total energy was converged to 1 meV/atom. Meanwhile, the Hellmann-Feymann force of the atom was converged to 0.001 eV/Å. All DFT calculations were spin polarized. Thus, the ferromagnetic (FM) ordering was assumed for Fe, Co and Ni elements. In the case of Cr, the anti-ferromagnetic (AFM) ordering was adopted. For other considered transition metals, the standard spin polarized calculations gave the non-paramagnetic (NM) states. The numerical integration grid points were carefully tested for meta-GGA functionals by means of varying the NGX, NGY and NGZ parameters in INCAR file of VASP inputs [27].

The four different semi-local density functionals employed in the calculations were Perdew-Zunger (LDA) [31], Perdew-Burke-Ernzerhof (PBE-GGA) [32], PBE revised for solids (PBEsol-GGA) [16], and r$^2$SCAN [33,35].

The elastic constants for all investigated crystal structures were obtained from VASP calculations directly using the strain energy density versus strain method with the help of VASPkit tool [34]. To calculate thermophysical properties such as heat capacity and volumetric thermal expansion coefficient as a function of temperature, the quasi-harmonic approximation (QHA) was employed for all structures. All QHA calculations were performed using the auxiliary tool in the Phonopy [35] and the in-house code, employing the $3\times3\times3$ (for cubic lattice structures) and $4\times4\times3$ (hexagonal crystal structures) supercells, resulting in the lattice parameter that is always larger than 9.5 Å in each crystallographic direction. The atomic forces were calculated for all displaced supercells using $4\times4\times4$ (or $4\times4\times3$ mesh for hexagonal lattice structures) k-mesh in VASP within the spin polarized DFT scheme, assuming the proper magnetic orderings for the considered transition metals, i.e., the FM (Fe, Co and Ni) and AFM (Cr) and NM states for others. The dependence of cell volume on temperature was obtained by fitting the free energy versus volume isothermal profiles to the Vinet equation of state (EOS) [35,36]. On each isothermal profile, the phonon spectra were calculated for nine different cell volumes, i.e., eight homogeneously deformed crystal structures (four volumes in either compressed or dilated conditions) and the equilibrium one. For metals, the electron thermal excitations at finite temperature also contribute to the entropy and free energy, which were considered in our calculations [1]. The equilibrium cell volume at 0 K was obtained by fitting the EOS to the total energy versus volume profile at the same temperature, with the inclusion of the zero-point vibrational energies for all sampled volumes in QHA methodology. The expressions for calculating the equilibrium lattice parameters, Debye temperature, volumetric thermal expansion coefficient and vibrational free energy were also provided in the supporting information as Eqs. (S1)-(S6).

Table I The equilibrium cell volumes (Å$^3$) of the 3$d$,4$d$ 5$d$ transition metals at 0 K, including zero-point energy corrections by different semilocal functionals. The experimental values data (Ref. [37][41]) are also given. Note that for all hexagonal crystal structure, the c/a ratio is determined by performing the standard DFT structural optimization procedure at 0 K.

| Phase | Space Group | LDA | PBE | PBEsol | r$^2$SCAN | Expt. |
|---|---|---|---|---|---|---|
| Ti | P 6$_3$/mmc | 15.93 | **17.13** | 16.55 | 17.03 | 17.50 [a] |
| V | Im -3 m | 12.62 | **13.44** | 12.97 | 13.32 | 13.78 [a] |
| Cr | Im -3 m | 10.85 | **11.53** | 11.13 | 11.36 | 11.82 [a] |
| Fe | Im -3 m | 10.39 | 11.40 | 10.81 | **11.77** | 11.64 [a] |
| Co | P 6$_3$/mmc | 9.97 | **10.82** | 10.35 | — | 10.96 [a] |
| Ni | F m -3 m | 9.87 | **10.68** | 10.21 | 10.42 | 10.81 [a] |
| Cu | F m -3 m | 10.95 | 10.95 | 11.36 | **11.49** | 11.65 [a] |
| Y | P 6$_3$/mmc | 29.82 | **32.79** | 31.25 | 33.50 | 32.95 [a] |
| Zr | P 6$_3$/mmc | 21.83 | **23.42** | 22.50 | 23.51 | 23.18 [a] |
| Nb | Im -3 m | 17.39 | 18.33 | **17.73** | 18.34 | 17.97 [a] |
| Mo | Im -3 m | 15.14 | 15.84 | **15.37** | 15.71 | 15.51 [a] |
| Ru | P 6$_3$/mmc | 13.13 | 13.77 | 13.33 | **13.56** | 13.45 [a] |
| Pd | F m -3 m | 14.17 | 15.26 | **14.52** | 14.98 | 14.56 [a] |
| Ag | F m -3 m | 16.06 | 17.84 | **16.66** | 17.32 | 16.85 [a] |
| Hf | P 6$_3$/mmc | 20.83 | **22.41** | 21.55 | 22.12 | 22.30 [a] |
| Ta | I m -3 m | 17.30 | 18.30 | 17.69 | **18.10** | 17.93 [a] |
| W | I m -3 m | 15.31 | 15.93 | 15.50 | **15.72** | 15.80 [a] |
| Re | P 6$_3$/mmc | 14.34 | 14.87 | 14.50 | **14.71** | 14.62 [a] |
| Os | P 6$_3$/mmc | 13.74 | 14.28 | **13.91** | 14.05 | 13.85 [a] |
| Ir | F m -3 m | 13.93 | 14.52 | **14.09** | 14.30 | 14.06 [a] |
| Pt | F m -3 m | 14.73 | 15.51 | **14.94** | 15.33 | 15.02 [a] |
| Au | F m -3 m | **16.65** | 17.97 | 17.02 | 17.59 | 16.82 [a] |
| **MAPE** | | 5.04% | 2.64% | 2.75% | 2.00% | |
| **MPE** | | -5.04% | 0.98% | -2.59% | 0.42% | |

[a] Ref. [37].

## 3. Results and Discussions

### 3.1. *Equilibrium cell volumes*

Using the QHA, the calculated equilibrium cell volumes of the considered 3*d*, 4*d* and 5*d* transition metals are given in Table I. Meanwhile, the calculated mean percentage errors are plotted in Figure 1 for equilibrium cell volumes. We should note that for hexagonal crystal structures, the *c/a* ratio is fixed to the value optimized by the corresponding exchange-correlation functional at 0 K before applying the QHA calculations. The fitting parameters of Vinet EOS for all investigated transition metals using different exchange-correlation functionals at 0 K are given in Tables. SII to SVI in the supporting information. All relevant data for Co are missing in current work for r$^2$SCAN method, because we could not eliminate the imaginary phonon modes of this particular element within a hexagonal lattice structure. Interestingly, using the same computational parameters and r$^2$SCAN method, imaginary phonon modes are not obtained for Co in the face-centered cubic form. The hexagonal Co lattice structure is predicted to be intrinsically unstable using r$^2$SCAN functional. The zero-point energy correction and QHA method cannot be applied to Co to obtain other thermophysical properties.

Here, the discrepancy between the theoretical value and that of experiment is calculated by the percentage error as PE=[(Cal-Exp)/Exp]×100%. In addition, the mean percentage error (MPE) and the mean absolute percentage error (MAPE) are obtained from Eqs. (1) and (2), respectively. Here, the *N* in Eqs. (1) and (2) refers to the total number of sampled data points. By the definition of MPE, the positive value indicates the overestimation of the targeted parameter by the theory. Otherwise, the negative value quantifies the degree of underestimation of the parameter by the calculation, compared to the experimental value. The overall performance of the method is characterized by the MAPE, i.e., the smaller MAPE deliveries the higher accuracy of the method.

$$\text{MPE} = \frac{1}{N}\sum_{i=1}^{N} PE_i \qquad (1)$$

$$\text{MAPE} = \frac{1}{N}\sum_{i=1}^{N} |PE_i| \qquad (2)$$

As can be seen from Figure. 1, the LDA functional tends to underestimate the equilibrium cell volume for most transition metals, as the calculated PE values are negative for most transition metals. Meanwhile, the opposite trend is found for PBE-GGA functional. Notably, the obtained PEs of PBEsol and r$^2$SCAN are placed in between those of PBE and LDA for the 3$d$, 4$d$ and 5$d$ transition metals. The MPEs are found to be -5.04% (LDA), +0.98% (PBE), -2.59% (PBEsol), and +0.42% (r$^2$SCAN), respectively.

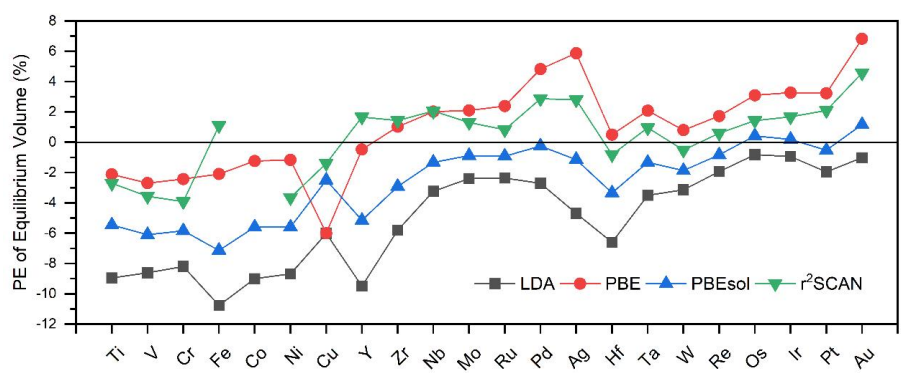

Figure 1. The obtained percentage errors of semilocal density functionals for equilibrium cell volumes of 22 transition metals.

Generally speaking, among the four tested LDA, GGA and meta-GGA density functionals (LDA, PBE, PBEsol, r$^2$SCAN), PBEsol and r$^2$SCAN could predict the equilibrium cell volume of transition metals in better agreement with experiments than those of either LDA or PBE-GGA [37]. By a close looking at the mean absolute percentage errors (MAPEs) of those functionals, the following conclusions could be drawn. First of all, r$^2$SCAN performs the best among all assessed semilocal functionals, and which gives the smallest MAPE as 2.00%. Meanwhile, the PBEsol performs much better than either LDA for the equilibrium cell volumes, compared to experimental values and those of r$^2$SCAN. The mean absolute percentage error (MAPE) for PBEsol is found to be 2.75%. For PBE and LDA, we obtain 2.64% (PBE) and 5.04% (LDA) for MAPEs. The outstanding performance of r$^2$SCAN functional on the equilibrium cell volume of various transition metals spanning from the 3$d$ to 5$d$ elements is clearly demonstrated in Table. 1, as the obtained MAPE values is the smallest among all tested semilocal functionals. Notably, for calculating equilibrium cell volumes of transition metals, PBEsol is found to be as promising as that of r²SCAN.

Table II The calculated cohesive energies (eV/atom) of the 3d,4d 5d transition metals at 0 K using different semilocal density functionals, corrected by zero-point energy. The experimental data (Ref. [37]) and other previous theoretical values of PBE (Ref. [38]) are given for comparisons.

| Structure | LDA | PBE | PBEsol | r²SCAN | Expt. | PBE (Ref.) |
|---|---|---|---|---|---|---|
| Ti | -6.55 | -5.66 | -6.10 | **-5.22** | -4.88[a] | -5.45[b] |
| V | -6.84 | **-5.49** | -6.05 | -4.96 | -5.34[a] | -6.03[b] |
| Cr | -5.72 | **-4.06** | -4.79 | -3.21 | -4.15[a] | -4.00[b] |
| Fe | -6.55 | -5.10 | -5.94 | **-4.76** | -4.32[a] | -4.87[b] |
| Co | -6.80 | -5.57 | -6.26 | — | -4.47[a] | -5.27[b] |
| Ni | -6.28 | -5.13 | -5.80 | **-4.71** | -4.48[a] | -4.87[b] |
| Cu | -4.54 | -4.50 | -4.05 | **-3.87** | -3.51[a] | -3.48[b] |
| Y | -4.93 | -4.34 | -4.77 | **-4.44** | -4.42[a] | -4.13[b] |
| Zr | -7.52 | **-6.51** | -7.00 | -6.06 | -6.32[a] | -6.16[b] |
| Nb | -8.51 | -6.98 | **-7.65** | -6.46 | -7.47[a] | -6.98[b] |
| Mo | -8.23 | -6.31 | **-7.12** | -5.64 | -6.84[a] | -6.21[b] |
| Ru | -8.99 | -7.22 | -8.15 | **-6.86** | -6.80[a] | -6.67[b] |
| Pd | -5.24 | **-3.89** | -4.62 | -4.17 | -3.93[a] | -3.71[b] |
| Ag | -3.64 | -2.52 | -3.09 | **-2.89** | -2.96[a] | -2.49[b] |
| Hf | -7.63 | **-6.75** | -7.29 | -7.33 | -6.44[a] | -6.40[b] |
| Ta | -9.70 | **-8.12** | -8.95 | -7.89 | -8.11[a] | -8.27[b] |
| W | -10.28 | -8.32 | **-9.11** | -8.06 | -8.83[a] | -9.07[b] |
| Re | -9.85 | **-7.87** | -8.85 | -7.50 | -8.06[a] | -7.82[b] |
| Os | -10.33 | **-8.47** | -9.49 | -8.69 | -8.22[a] | -8.29[b] |
| Ir | -9.31 | **-7.65** | -8.64 | -8.18 | -6.96[a] | -7.32[b] |
| Pt | -7.17 | **-5.67** | -6.52 | -6.05 | -5.87[a] | -5.50[b] |
| Au | -4.30 | -3.03 | **-3.70** | -3.41 | -3.83[a] | -2.99[b] |
| MAPE | 27.14% | 8.99% | 15.17% | 8.42% | | |
| MPE | 27.14% | 2.96% | 14.86% | -0.35% | | |

[a] Ref. [37].
[b] Ref. [38].

*3.2 Cohesive energy*

The cohesive energy of a transition metal is obtained as the energy difference between the bulk crystal structure and the constituting atoms. In order to reliably predict the cohesive energy, the employed density functional must be able to calculate the total energy of crystal structure as well as that of an isolated atom with high accuracy concomitantly.

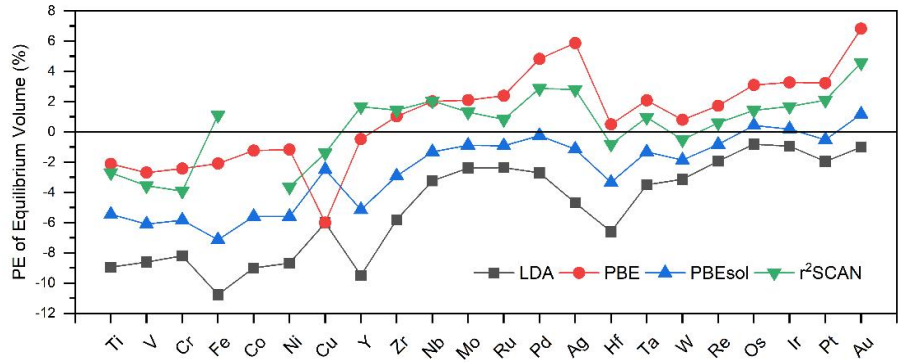

Figure 2. The obtained percentage errors of cohesive energy (eV/atom) of 22 transition metals using semilocal functionals.

    The cohesive energies of the transition metals are listed in Table. II using different density functionals together with experimental values. The calculated percentage errors are illustrated in Figure. 2 for each elemental crystal structure. First, we notice that LDA, which gives a large positive PEs of +27.14%, overestimates the sizes of cohesive energy of all investigated transition metals, i.e., the calculated cohesive energy by LDA is significantly more negative than experimental values. PBE tends to overestimate the cohesive energies for 3$d$ transition metals (Ti, V, Cr, Fe, Co, Ni and Cu), similar to LDA, but with a significantly reduced PE for each case. For most 4$d$ and 5$d$ transition metals, PBE performs well for the cohesive energies. In the cases of Ag and Au, PBE strongly underestimates the binding strength of bulk crystals, leading to large negative PEs. It is worth mentioning that PBE also greatly overestimates equilibrium cell volumes of Ag and Au at the same time, as shown in Table I. The behavior of PBEsol for calculating cohesive energies of 3$d$-to-5$d$ transition metals closely follows that of LDA rather than those of PBE except that of Au. Among all tested 3$d$, 4$d$, and 5$d$ transition metals, PBEsol underestimates the size of cohesive energy of Au.   Interestingly, PBEsol does not follow LDA for predicting the equilibrium cell volumes of transition metals, as shown in Figure. 1. Therefore, the strong over binding of solids by PBEsol is due to the poor accuracy for calculating the total energies of isolated atoms. For predicting the cohesive energies of various transition metals, the best performance is obtained from r²SCAN. As shown in Figure. 2, r²SCAN decreases the PEs for most 3$d$ transition metals, compared to PBE functionals. Meanwhile, r²SCAN also as expected performs very well for predicting

the cohesive energies of 4$d$ and 5$d$ elements with the more delocalized valence orbitals.

Finally, the calculated MPEs and MAPEs are given for the tested density functionals in Table. 2. LDA and PBEsol give very large values for both quantities, i.e., LDA (MPE: 27.14%; MAPE: 27.14%) and PBEsol (MPE: 14.86%; MAPE: 15.17%). The strong overbinding of solids predicted by either LDA or PBEsol is clearly seen from the obtained positive value of MPE. The r$^2$SCAN is slightly better than PBE for the cohesive energy of transition metals with a smaller MAPE (8.42%) than that of PBE (8.99%).

*3.3 Bulk modulus*

The bulk modulus is obtained as one of the fitting parameters in the EOS, and practically represents the second derivative of total cell energy with respect to the cell volume, i.e., the curvature of the energy versus volume profile. Thus, the bulk modulus tests the ability of an exchange-correlation functional to reliably describe the variation of total cell energy versus elastic structural deformation. Experimentally, the bulk modulus can be routinely measured using the ultrasonic resonance method for single or multi-crystalline bulk materials. Here, we would like to emphasize the possible large uncertainty involved in the experimental values of bulk modulus for transition metals due to influences of the factors such as microstructures, crystallinity, defects, and temperature on the measurement.

The calculated bulk moduli of transition metals by using the four different exchange-correlation functionals and the associated MPEs and MAPEs are given in Table. III. At the same time, the obtained percentage errors are also displayed in Figure. 3. Obviously, LDA performs worse than other tested methods, and tends to overestimate the bulk modulus of transition metals regardless of the degree of the localization of $d$ states. As a result, the obtained PEs of LDA for all transition metals are positive. Such behavior is highly consistent with the fact that LDA underestimates the equilibrium cell volumes and strongly overbinds the solids, as can be seen from Figures. 1 and 2. The performance of PBEsol for many transition metals closely resembles that of LDA, but the PEs are significantly reduced, indicating that the

PBEsol predicts the bulk modulus of transition metals with a higher accuracy than LDA. For the tested GGA functionals, PBE provides even better results for bulk modulus of transition metals than PBEsol except few cases including Fe, Ru, Pd, Pt, Ag, Os, and Au. Specifically, PBE does give relatively substantial percentage errors in predicting the bulk modulus of elemental Cr (+26.2%), Cu (+28.6%) or Au (-23.59%). In the case of Au, the large error in bulk modulus using PBE functional is attributed to the strongly overestimation of the equilibrium cell volumes (See Figure. 1). Figure. 3 also shows that r²SCAN performs better than PBE for computing the bulk modulus of 3$d$, 4$d$ and 5$d$ transition metals in general.

Table III The calculated bulk modulus (GPa) of the 3$d$,4$d$ 5$d$ transition metals at 0 K with the zero-point energy corrections to EOS curve using semilocal functionals. The results are compared to experimental data (Ref. [37,39]) and previous theoretical values of PBE (Ref. [38]) functional.

| Structure | LDA | PBE | PBEsol | r²SCAN | Expt | PBE (Ref.) |
|---|---|---|---|---|---|---|
| Ti | 133.0 | **116.6** | 125.1 | 123.6 | 108.3[a] | 113.5[c] |
| V | 227.4 | **193.8** | 211.5 | **193.8** | 165.8[b] | 183.1[c] |
| Cr | 302.1 | **258.3** | 283.0 | 275.8 | 204.6[b] | 261.2[c] |
| Fe | 255.7 | **185.2** | 218.8 | 201.4 | 175.0[b] | 195.3[c] |
| Co | 268.3 | **214.3** | 246.8 | — | 198.4[b] | 212.5[c] |
| Ni | 264.4 | **210.0** | 240.8 | 226.4 | 192.5[b] | 193.9[c] |
| Cu | 183.0 | 185.5 | 168.7 | **162.8** | 144.3[b] | 146.9[c] |
| Y | 43.9 | 41.0 | **42.3** | 40.9 | 41.7[a] | 40.7[c] |
| Zr | 103.3 | 93.6 | 98.9 | **95.4** | 95.9[a] | 95.5[c] |
| Nb | 194.9 | **173.4** | 186.6 | 173.7 | 173.2[b] | 171.1[c] |
| Mo | 289.6 | 259.4 | 279.1 | 275.0 | 276.2[b] | 261.3[c] |
| Ru | 365.8 | 311.9 | 347.8 | **337.5** | 335.5[b] | 308.2[c] |
| Pd | 229.6 | 172.8 | 207.4 | **184.6** | 195.4[a] | 169.4[c] |
| Ag | 136.9 | 91.0 | 118.3 | **104.4** | 105.7[b] | 83.3[c] |
| Hf | 121.6 | **113.9** | 120.1 | 117.4 | 110.7[b] | 108.0[c] |
| Ta | 219.7 | 197.4 | 208.8 | **205.7** | 202.7[b] | 195.3[c] |
| W | 345.6 | 309.5 | 333.7 | **331.2** | 327.5[b] | 316.2[c] |
| Re | 414.2 | **373.7** | 401.8 | 393.8 | 380.8[b] | 372.1[c] |
| Os | 453.0 | 397.6 | 434.9 | **434.7** | 424.6[b] | 402.6[c] |
| Ir | 403.6 | 348.6 | 388.7 | **377.4** | 365.2[b] | 347.3[c] |
| Pt | 316.0 | 263.0 | 301.2 | **272.3** | 285.5[b] | 250.9[c] |
| Au | 191.3 | 139.1 | **174.4** | 153.5 | 182.0[b] | 138.4[c] |
| MAPE | 18.41% | 9.10% | 11.14% | 7.62% | | |
| MPE | 18.41% | 0.47% | 10.76% | 4.87% | | |

[a] Ref. [39].

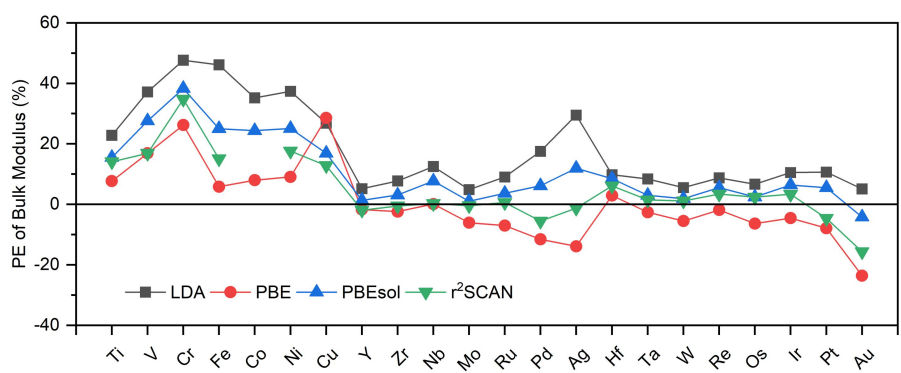

Figure 3. The calculated percentage errors between the theoretical and experimental values of bulk moduli of 22 transition metals by semilocal functionals.

The calculated MPEs are +18.41% (LDA), +0.47% (PBE), +10.76% (PBEsol), and +4.87% (r$^2$SCAN), respectively. Therefore, the bulk modulus of 3$d$, 4$d$ and 5$d$ transition metals is overestimated by most semilocal functionals except PBE. Meanwhile, the obtained MAPEs illustrate the promising performance of r$^2$SCAN and PBE for calculating the bulk modulus of transition metals. Nevertheless, r$^2$SCAN gives the smallest MAPE (7.62%) among all four methods for bulk modulus.

*3.4 Thermophysical properties*

Regarding the thermophysical properties, we would like to address the Debye temperature and volumetric thermal expansion coefficients. Those two properties are particularly discussed not only because both are closely relevant to the engineering applications of transition metals and their alloys at finite temperature[40-42], but also due to the availability of the benchmark experimental values for the comparisons. Similar to the case of bulk modulus, large uncertainties are also expected in the experimental values of Debye temperature and thermal expansion coefficient. Therefore, the results presented here are mainly served to provide an overview of the state-of-the-art semilocal exchange-correlation functionals in calculating thermophysical parameters at finite temperature, specifically assessing the feasibility and numerical stability of applying r$^2$SCAN functional for the relevant calculations.

Table IV The calculated Debye temperature (K) of the 3*d*,4*d* 5*d* transition metals, compared with the experimental data (Ref.[40][41]).

| Phase | LDA | PBE | PBEsol | r$^2$SCAN | Expt |
|---|---|---|---|---|---|
| Ti | 315 | 390 | 386 | **420** | 420[a] |
| V | 285 | **365** | 361 | 298 | 380[a] |
| Cr | 616 | 606 | **624** | 586 | 630[a] |
| Fe | 547 | **464** | 515 | 429 | 470[a] |
| Co | 567 | **516** | 549 | — | 460[a] |
| Ni | 538 | **499** | 529 | 516 | 477[a] |
| Cu | 298 | **326** | 283 | 371 | 346[a] |
| Y | 241 | 250 | **251** | 244 | 300[b] |
| Zr | 233 | 270 | 258 | **295** | 299[a] |
| Nb | 227 | 250 | **251** | 222 | 277[a] |
| Mo | 457 | 458 | **464** | 484 | 470[a] |
| Ru | 582 | **550** | 574 | 565 | 555[a] |
| Pd | 297 | **264** | 288 | 256 | 275[a] |
| Ag | **228** | 199 | 219 | 204 | 224[a] |
| Hf | 236 | **245** | 243 | 263 | 252[a] |
| Ta | 257 | **258** | 263 | 257 | 258[a] |
| W | 392 | **380** | 393 | 383 | 380[a] |
| Re | **425** | 396 | 486 | 390 | 416[a] |
| Os | 488 | **469** | 486 | 483 | 467[a] |
| Ir | 451 | **427** | 448 | 442 | 425[a] |
| Pt | 248 | 225 | **244** | 217 | 237[a] |
| Au | **156** | 114 | 147 | 107 | 163[a] |
| MAPE | 10.35% | 6.03% | 7.58% | 8.09% | |
| MPE | -2.14% | -4.40% | -0.01% | -4.95% | |

[a] Ref. [40].

[b] Ref. [41].

The predicted Debye temperatures and percentage errors of the four semilocal functionals for each transition metal are shown in Table IV and Figure. 4, respectively. As can be seen from Figure. 4, all tested semilocal functionals provide similar accuracies for computing Debye temperature for most transition metals. The Debye temperature of a solid is usually calculated from the equilibrium lattice volume (mass density) and bulk modulus (sound velocity). Therefore, the errors in the obtained Debye temperature for different exchange-correlation functionals are directly inherited from the accuracies in evaluating the equilibrium cell volumes and bulk modulus. However, the errors in cell volumes and bulk modulus usually counteract with each other. Therefore, the calculation of Debye temperature may not sensitive to the employed exchange-correlation functional. Nevertheless, the MPEs are found to be -2.14% (LDA), -4.40% (PBE), -0.01% (PBEsol), and -4.95% ($r^2$SCAN), respectively. We may conclude that PBE and $r^2$SCAN tend to underestimate the Debye temperature for most investigated transition metals. The obtained MPEs increase in the order of PBEsol, LDA, PBE and $r^2$SCAN. Certainly, the results are in great contrast to the high accuracy of $r^2$SCAN for calculating the equilibrium cell volumes and bulk modulus for transition metals, compared to those of LDA and PBE. Regarding the experimental values of Debye temperature for transition metals, the numbers are scattered for each element in different literatures, and large uncertainties in the selected benchmark data should be kept in mind before drawing the conclusion.

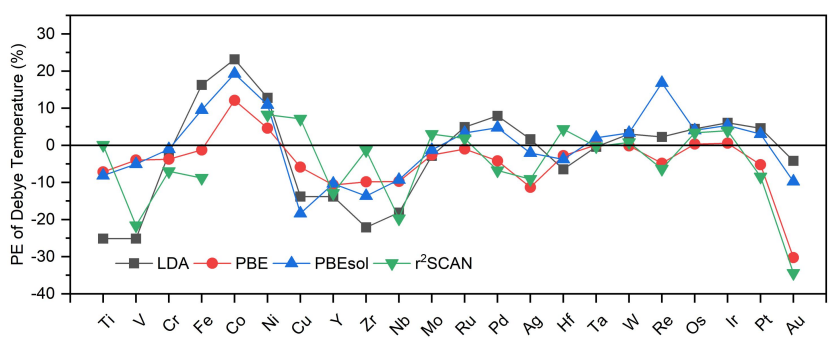

Figure 4. The obtained percentage errors between the theoretical and experimental values of Debye temperature (K) of 22 transition metals using different semilocal functionals.

The volumetric thermal expansion coefficient (TEC) versus temperature profiles for all studied 3$d$, 4$d$ and 5$d$ transition metals are displayed in Figure. SI in supporting materials. In Table. V, the average TEC, which is calculated in a temperature range from 300 K to 1000 K, is given for each transition metal using four different semilocal functionals. Unlike Debye temperature, the equilibrium cell volumes play the decisive role in determining the volumetric TEC for crystalline materials. Overestimation of cell volumes of a crystal structure by an employed exchange-correlation functional is anticipated to fatally deteriorate its reliability in predicting the thermal dilation at finite temperatures. In Figure. 5, the calculated PEs of semilocal functionals for TEC are illustrated. First of all, we observe a systematic overestimation of volumetric TECs of transition metals using PBE functional, as the obtained MPE and MAPE are +8.34% and 20.15%, respectively. Otherwise, LDA tends to underestimate the thermal dilation for most investigated transition metals, as the obtained MPE and MAPE are -8.10% and 19.60%, respectively. Meanwhile, r$^2$SCAN performs similarly to PBEsol for calculating volumetric TECs in terms of MPE and MAPE. Specifically, we find r$^2$SCAN (MPE: +4.06%; MAPE: 18.15%), compared to PBEsol (MPE: -4.87%; MAPE: 13.37%). Overall, PBEsol performs the best for thermal expansion coefficient, followed by r$^2$SCAN.

Table V The calculated average volumetric thermal expansion coefficient (10$^{-5}$ K$^{-1}$) of the 3$d$, 4$d$ 5$d$ transition metals in the temperature range from 300 K to 1000 K using semilocal functionals, and the comparison with experimental data (Ref. [42]).

| Phase | LDA | PBE | PBEsol | r$^2$SCAN | Expt |
|---|---|---|---|---|---|
| Ti | 2.5 | **3.0** | 2.8 | 2.7 | 3.2[a] |
| V | **3.0** | 2.7 | 2.8 | 3.6 | 3.1[a] |
| Cr | 2.1 | 2.3 | **2.4** | 1.7 | 3.0[a] |
| Fe | 2.9 | **4.0** | 2.6 | 3.6 | 4.6[a] |
| Co | 2.7 | **4.2** | 3.4 | — | 4.4[a] |
| Ni | 3.4 | 4.2 | 3.1 | **4.5** | 4.9[a] |
| Cu | 8.8 | 4.3 | 4.9 | **5.7** | 5.7[a] |
| Y | 2.1 | 2.7 | **3.1** | 2.4 | 3.7[a] |
| Zr | 1.5 | **2.3** | 2.4 | 2.0 | 2.2[a] |
| Nb | 2.9 | 2.9 | **2.8** | 3.5 | 2.4[a] |
| Mo | **1.6** | 2.0 | 1.7 | 1.8 | 1.6[a] |
| Ru | 2.0 | **2.4** | 2.1 | **2.2** | 2.3[a] |
| Pd | 3.6 | 4.6 | **4.0** | 4.5 | 4.2[a] |
| Ag | 5.7 | 11.8 | **7.2** | 8.9 | 6.8[a] |
| Hf | 1.9 | 2.4 | 2.1 | **2.0** | 2.0[a] |
| Ta | 2.7 | **2.3** | 2.4 | 1.9 | 2.1[a] |
| W | 1.4 | 1.6 | 1.5 | **1.4** | 1.4[a] |
| Re | **2.0** | 2.1 | 1.9 | 2.1 | 2.0[a] |
| Os | 1.6 | 1.8 | **1.7** | 1.7 | 1.7[a] |
| Ir | 2.1 | 2.5 | 2.2 | **2.3** | 2.3[a] |
| Pt | 2.9 | 3.7 | **3.1** | 3.9 | 3.0[a] |
| Au | **5.6** | 9.3 | 6.4 | 9.1 | 4.9[a] |
| MAPE | 19.60% | 20.15% | <span style="color:red">13.37%</span> | 18.15% | |
| MPE | -8.10% | 8.34% | -4.87% | 4.06% | |

[a] Ref. [42].

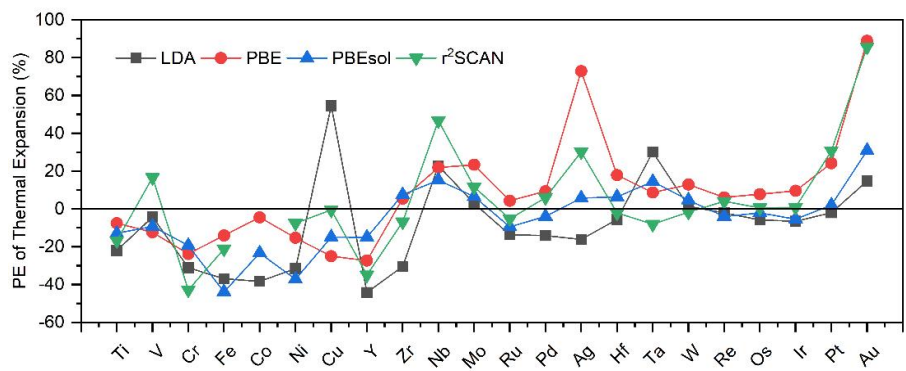

Figure 5. The calculated percentage errors between the theoretical and experimental values of the average volumetric thermal expansion coefficients of 22 transition metals for the temperature in a range from 300 K to 1000 K using five different semilocal functionals and QHA method.

Finally, we summarize our main findings in Figure. 6 for various computed properties of $3d$, $4d$ and $5d$ transition metals using LDA, PBE, PBEsol, and r$^2$SCAN functionals. Here, we divide the transition metals into two distinct groups, the $3d$ and $4d+5d$ categories. $3d$ transition metals possess more localized $3d$ states and the magnetic ordering is quite often seen in the ground state electronic structure. Otherwise, $4d+5d$ group has more delocalized $d$ states with the paramagnetism as the ground state in bulk crystal structures.

For equilibrium cell volumes (Figure. 6(a)) and bulk modulus (Figure. 6(b)) of $3d$ transition metals, PBE is able to deliver the same level of accuracy to that of r$^2$SCAN. Both r$^2$SCAN and PBE perform much better than LDA and PBEsol for the two properties of $3d$ elements. For the $4d+5d$ group, PBEsol and r$^2$SCAN are the better options for calculating cell volumes and bulk modulus than either LDA or PBE. For all considered $3d$, $4d$, and $5d$ transition metal elements, the best performance is achieved by r$^2$SCAN for the two properties.

Regarding the cohesive energy, as shown in Figure. 6(c), r$^2$SCAN offers the best performance among all tested semilocal functionals regardless of the different categories of transition metals. Besides r$^2$SCAN, PBE functional is better than PBEsol for computing cohesive energy of transition metals. LDA is the least accurate method for such calculations among all tested semilocal functionals.

For Debye temperature or volumetric TECs, the obtained MAPEs for different exchange-correlation functionals are given in Figure. S2(a) and (b). It is worth noting that for those properties, reliable experimental values are usually difficult to obtain. The measured results can be scattered in a wide range from different literatures for Debye temperature and thermal expansion coefficients. The comparisons and discussions of the observed trends in MPEs are omitted here to avoid the possible misinterpretation of the results.

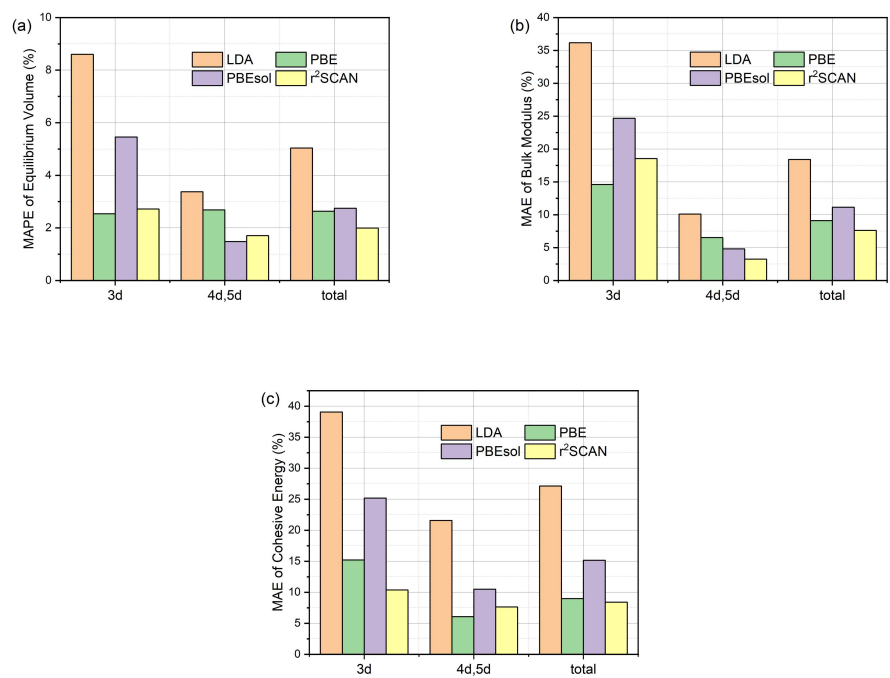

Figure. 6 The mean percentage errors of cell volume (a) bulk modulus (b) and cohesive energy (c) of transition metals, obtained from semilocal functionals.

## 4. Conclusions

Our calculations demonstrated that the r$^2$SCAN functional deliveries the well-balanced performances among all tested density functionals (LDA, PBE, PBEsol and r$^2$SCAN) for equilibrium cell volumes, cohesive energy, mechanical modulus, Debye temperature, and thermophysical properties of 3$d$, 4$d$ and 5$d$ transition metals. PBEsol offers the best agreement with the experiments for predicting the thermal expansion coefficients, while PBE performs best for predicting Debye temperature. Otherwise, for magnetic 3$d$ transition metals, PBE offers the best performance for calculating the equilibrium cell volumes, bulk modulus, and cohesive energy. The high accuracy and the sound numerical efficiency enable the use of r$^2$SCAN functional as the workhorse method in the future high throughput computational study of high-temperature super-alloys and high entropy alloys.

**Conflicts of interest**

There are no conflicts of interest to declare.

**Data Availability**

The data used to support the findings of this study are available from the corresponding author upon request.

**Supplementary materials**

See supplementary materials for the VASP potential file, crystal volume, volumetric thermal expansion coefficient (TEC) versus, and QHA fitting parameters for fixed geometries.

**Acknowledgements**

Bing Xiao would also like to acknowledge the financial support from the "Young Talent Supporting Program" of Xi'an Jiaotong University (No. DQ1J009). J.N., A.R, R.Z., and J.S. acknowledge the support of the U.S. Office of Naval Research (ONR) Grant No. N00014-22-1-2673. This work at Tulane University also used Purdue Anvil CPU at Rosen Center for Advanced Computing (RCAC) through allocation DMR190076 from the Advanced Cyberinfrastructure Coordination Ecosystem: Services & Support (ACCESS) program, which is supported by National Science Foundation grants #2138259, #2138286, #2138307, #2137603, and #2138296.

**References**

[1] M. B. Henderson, D. Arrell, R. Larsson, M. Heobel and G. Marchant, 'Nickel based superalloy welding practices for industrial gas turbine applications' Sci. Technol. Weld. Join. 9(1): 13-21, (2004). https://doi.org/10.1179/136217104225017099

[2] A.M. Mateo Garcia, 'BLISK fabrication by linear friction welding', Adv. Gas Turbine Technol. (2011).

[3] L. Zhu, N. Li, P. R. N. Childs, 'Light-weighting in aerospace component and system design' Propuls. Power Res. 7 (2) (2018). https://doi.org/10.1016/j.jppr.2018.04.001

[4] T. Wu, M. Sun, H. Wong, B. Huang 'Decoding of crystal synthesis of fcc-hcp reversible transition for metals: theoretical mechanistic study from facet control to phase transition engineering' Nano Energy,85(106026):1-12, 2021. https://doi.org/10.1016/j.nanoen.2021.106026

[5] P. H. T. Philipsen, E. J. Baerends, 'Cohesive energy of 3d transition metals: Density functional theory atomic and bulk calculations', Physical Review B, 54(8), 5326 (1996). https://doi.org/10.1103/PhysRevB.54.5326

[6] N. K. Nepal, S. Adhikari, J. E. Bates, A. Ruzsinszky, 'Treating different bonding situations: Revisiting Au-Cu alloys using the random phase approximation', Physical Review B, 100, 045135 (2019). https://doi.org/10.1103/PhysRevB.100.045135

[7] J. P. Perdew; A. Ruzsinszky; J. Tao; V. N. Staroverov; G. E. Scuseria; G. I. Csonka, 'Prescription for the design and selection of density functional approximations: More constraint satisfaction with fewer fits' J. Chem. Phys. 123, 062201 (2005). https://doi.org/10.1063/1.1904565

[8] G. I. Csonka, J. P. Perdew, A. Ruzsinszky, P. H. T. Philipsen, S. Lebègue, J. Paier, O. A. Vydrov, J. G. Ángyán, 'Assessing the performance of recent density functionals for bulk solids', Phys. Rev. B, 79, 155107 (2009) https://doi.org/10.1103/PhysRevB.79.155107

[9] F. Tran; J. Stelzl; P.r Blaha, 'Rungs 1 to 4 of DFT Jacob's ladder: Extensive test on the lattice constant, bulk modulus, and cohesive energy of solids', J. Chem. Phys, 144, 204120 (2016). https://doi.org/10.1063/1.4948636

[10] G. I. Csonka, J. P. Perdew, A. Ruzsinszky, P. H. T. Philipsen, S. Lebègue, J. Paier, O. A. Vydrov, J. G. Ángyán, 'Assessing the performance of recent density functionals for bulk solids' Phys. Rev. B, 79, 155107 (2009) https://doi.org/10.1103/PhysRevB.79.155107

[11] M. Ernzerhof; G. E. Scuseria, 'Assessment of the Perdew–Burke–Ernzerhof exchange-correlation functional', Chem. Phys, 110 (11), 5029 (1999). https://doi.org/10.1063/1.478401

[12] P. Haas, F. Tran, P. Blaha, K. Schwarz, R. Laskowski, 'Insight into the performance of GGA functionals for solid-state calculations', Phys. Rev. B, 80, 195109 (2009). https://doi.org/10.1103/PhysRevB.80.195109

[13] B. Xiao, J. Sun, A. Ruzsinszky, J. Feng, R. Haunschild, G. E. Scuseria, J. P. Perdew, 'Testing

density functionals for structural phase transitions of solids under pressure: Si, $SiO_2$, and Zr', Phys. Rev. B, 88, 184103 (2013). https://doi.org/10.1103/PhysRevB.88.184103

[14] J. Sun, A. Ruzsinszky, J. P. Perdew, 'Strongly Constrained and Appropriately Normed Semilocal Density Functional', Phys. Rev. Lett, 115, 036402 (2015). https://doi.org/10.1103/PhysRevLett.115.036402

[15] J. Sun, R. C. Remsing, Y. Zhang, Z. Sun, A. Ruzsinszky, H. Peng, Z. Yang, A. Paul, U. Waghmare, X. Wu, M. L. Klein, J. P. Perdew, 'Accurate first-principles structures and energies of diversely bonded systems from an efficient density functional', Nat. Chem., 8: 831-836 (2016) https://doi.org/10.1038/nchem.2535

[16] J. P. Perdew, K. Burke, M. Ernzerhof, 'Generalized Gradient Approximation Made Simple', Phys. Rev. Lett. 77, 3865 (1996). https://doi.org/10.1103/PhysRevLett.77.3865

[17] Y. Zhang, D. A. Kitchaev, J. Yang, T. Chen, S. T. Dacek, R. A. Sarmiento-Pérez, M. A. L. Marques, H. Peng, G. Ceder, J. P. Perdew, J. Sun, 'Efficient first-principles prediction of solid stability: Towards chemical accuracy', Npj Comput. Mater. 4, 9 (2018). https://doi.org/10.1038/s41524-018-0065-z

[18] D. A. Kitchaev, H. Peng, Y. Liu, J. Sun, J. P. Perdew, and G. Ceder, 'Energetics of $MnO_2$ polymorphs in density functional theory', Phys. Rev. B 93, 045132 (2016). https://doi.org/10.1103/PhysRevB.93.045132

[19] J. H. Yang, D. A. Kitchaev, G. Ceder, 'Rationalizing accurate structure prediction in the meta-GGA SCAN functional', Phys. Rev. B 100, 035132 (2019). https://doi.org/10.1103/PhysRevB.100.035132

[20] E. B. Isaacs, C. Wolverton, 'Performance of the strongly constrained and appropriately normed density functional for solid-state materials', Phys. Rev. Mat. 2, 063801 (2018). https://doi.org/10.1103/PhysRevMaterials.2.063801

[21] R. Sundararaman, K. Letchworth-Weaver. K. A. Schwarz, 'Improving accuracy of electrochemical capacitance and solvation energetics in first-principles calculations', J. Chem. Phys.149, 144105 (2018). https://doi.org/10.1063/1.5024219

[22] S. Jana, K. Sharma, P. Samal, 'Assessing the performance of the recent meta-GGA density functionals for describing the lattice constants, bulk moduli, and cohesive energies of alkali, alkaline-earth, and transition metals', J. Chem. Phys.149, 164703 (2018).

https://doi.org/10.1063/1.5047863

[23] M. Ekholm, D. Gambino, H. J. M. Jönsson, F. Tasnádi, B. Alling, I. A. Abrikosov, 'Assessing the SCAN functional for itinerant electron ferromagnets', Phys. Rev. B, 98, 094413 (2018). https://doi.org/10.1103/PhysRevB.98.094413

[24] Y. Zhang, W. Zhang, D. J. Singh, 'Localization in the SCAN meta-generalized gradient approximation functional leading to broken symmetry ground states for graphene and benzene', Phys. Chem.Chem. Phys., 22, 19585-19591 (2020). https://doi.org/10.1039/D0CP03567J

[25] Y, Fu and D, J. Singh, 'Applicability of the Strongly Constrained and Appropriately Normed Density Functional to Transition-Metal Magnetism', Phys. Rev. Lett. 121, 207201 (2018) https://doi.org/10.1103/PhysRevLett.121.207201

[26] A. P. Bartók, J. R. Yates, 'Regularized SCAN functional', J. Chem. Phys.150, 161101 (2019) https://doi.org/10.1063/1.5094646

[27] J. W. Furness, A. D. Kaplan, J. Ning, J. P. Perdew, J. Sun*, 'Accurate and Numerically Efficient r$^2$SCAN Meta-Generalized Gradient Approximation', J. Phys. Chem. Lett., 11, 8208−8215. (2020). https://doi.org/10.1021/acs.jpclett.0c02405

[28] G. Kresse and J. Furthmüller, 'Efficient iterative schemes for ab initio total-energy calculations using a plane-wave basis set', Phys. Rev. B. 54, 11169–11186. (1996) https://doi.org/10.1103/PhysRevB.54.11169

[29] G. Kresse and J. Hafner, 'Ab initio molecular-dynamics simulation of the liquid-metal–amorphous-semiconductor transition in germanium', Phys. Rev. B, 49, 14251 (1994). https://doi.org/10.1103/PhysRevB.49.14251

[30] J. D. Pack and H. J. Monkhorst, '"Special points for Brillouin-zone integrations"—a reply', Phys Rev B,16:1748. (1976) https://doi.org/10.1103/PhysRevB.16.1748

[31] P.E. Blochl, 'Projector augmented-wave method', Phys. Rev. B, 50, 17953 (1994). https://doi.org/10.1103/PhysRevB.50.17953

[32] J. P. Perdew and A. Zunger, 'Self-interaction correction to density-functional approximations for many-electron systems', Phys. Rev. B 23, 5048 (1981). https://doi.org/10.1103/PhysRevB.23.5048

[33] J. P. Perdew, A. Ruzsinszky, G. I. Csonka, O. A. Vydrov, G. E. Scuseria, L. A. Constantin, X.

Zhou, and K. Burke, 'Restoring the Density-Gradient Expansion for Exchange in Solids and Surfaces', Phys. Rev. Lett. 100, 136406 (2008).

https://doi.org/10.1103/PhysRevLett.100.136406

[34] V. Wang, N. Xu, J.C. Liu, G. Tang, W.T. Geng, 'VASPKIT: A User-Friendly Interface Facilitating High-Throughput Computing and Analysis Using VASP Code',Comput. Phys. Commun. 267, 108033 (2021). https://doi.org/10.1016/j.cpc.2021.108033

[35] A. Togo and I. Tanaka, 'First principles phonon calculations in materials science', Scr. Mater. 108, 1 (2015). https://doi.org/10.1016/j.scriptamat.2015.07.021

[36] Y. Wang, L. G. Hector, H. Zhang, S. L. Shang, L.Q. Chen, Z. K. Liu, 'Thermodynamics of the Ce γ–α transition: Density-functional study', Phys. Rev. B 78 104113. (2008) https://doi.org/10.1103/PhysRevB.78.104113

[37] K. Lejaeghere, V. V. Speybroeck, G. V. Oost, S. Cottenier, 'Error Estimates for Solid-State Density-Functional Theory Predictions: An Overview by Means of the Ground-State Elemental Crystals', Crit. Rev. Solid State Mater. Sci. 39, 1. (2014) https://doi.org/10.1080/10408436.2013.772503

[38] P. Janthon, S. Luo, S. M. Kozlov, F. Viñes, J. Limtrakul, D. G. Truhlar, F. Illas, 'Bulk Properties of Transition Metals: A Challenge for the Design of Universal Density Functionals', J. Chem. Theory Comput., 10, 9, 3832–3839. (2014) https://doi.org/10.1021/ct500532v

[39] A. Patra, J. E. Bates, J. Sun, and J. P. Perdew, 'Properties of real metallic surfaces: Effects of density functional semilocality and van der Waals nonlocality', Proc. Natl. Acad. Sci. U. S. A. 114(44), E9188–E9196 (2017). https://doi.org/10.1073/pnas.1713320114

[40] M. A. Meyers. 'Mechanical Behavior of Materials'. Cambrige University Press:New York,(2008). https://doi.org/10.1017/CBO9780511810947

[41] C. Kittel, 'Introduction to Solid State Physics, 8th edition'. John Wiley & Sons, Inc, (2005).

[42] Y. S. Touloukian, R. K. Kirby, R. E. Taylor, P. D. Desai. 'Thermal Expansion: Metallic Elements and Alloys', Thermophysical Properties of Matter, 12 (1975).

Supporting Information

# "Assessing r$^2$SCAN meta-GGA functional for structural parameters, cohesive energy, mechanical modulus and thermophysical properties of 3*d*, 4*d* and 5*d* transition metals"

Haoliang Liu[a], Xue Bai[b], Jingliang Ning[c], Yuxuan Hou[a], Zifeng Song[a], Akilan Ramasamy[c], Ruiqi Zhang[c], Yefei Li[b*], Jianwei Sun[c*], Bing Xiao[a*]

[a] *State Key Laboratory of Electric Insulation and Power Equipment and School of Electrical Engineering, Xi'an Jiaotong University, Xi'an 710049, China*
[b] *State Key Laboratory for Mechanical Behavior of Materials, Xi'an Jiaotong University, Xi'an 710049, China*
[c] *Department of Physics and Engineering Physics, Tulane University, New Orleans, , United States*

\* Correspondence authors: Yefei Li, E-mail: liyefei@xjtu.edu.cn; Jianwei Sun, E-mail: jsun@tulane.edu; Bing Xiao, E-mail: bingxiao84@xjtu.edu.cn

**Relevant expressions for calculating thermophysical properties**

The expression of equation of state is given by Eq. (S1) [1].

$$E(V) = E_0 + \frac{9V_0 B_0}{16}\left\{\left[\left(\frac{V_0}{V}\right)^{2/3} - 1\right]^3 B_0' + \left[\left(\frac{V_0}{V}\right)^{2/3} - 1\right]^2 \left[6 - 4\left(\frac{V_0}{V}\right)^{2/3}\right]\right\} \quad (S1)$$

In this equation, $E_0$ is the energy of equilibrium state, $V_0$ is the volume of equilibrium state; $B_0$ is the bulk modulus, $B_0'$ is the derivative of the bulk modulus with respect to pressure. Fitting Eq. (S1) to the energy versus volume data obtained from QHA calculations, the equilibrium cell volume and bulk modulus are calculated.

The Debye temperature can be described by the Eqs. (S2)-(S4).

$$\Theta = \frac{h}{k}\left[\frac{3n}{4\pi}\left(\frac{N_A \rho}{M}\right)\right]^{1/3} v_m \quad (S2)$$

$$v_m = \left[\frac{1}{3}\left(\frac{2}{v_l^3} + \frac{1}{v_t^3}\right)\right]^{-1/3} \quad (S3)$$

$$\begin{cases} v_l = \sqrt{(B + \frac{4}{3}G)/\rho} \\ v_t = \sqrt{\frac{G}{\rho}} \end{cases} \quad (S4)$$

In this formula, $\Theta$ is Debye temperature, $\hbar = h/2\pi$ and $h$ is Planck constant, $k$ is Boltzmann constant, and $N_A$ represents the Avogadro's number; $\rho$ is the density of the unite cell; n denotes the number of atoms per chemical formula; $M$ refers to the molecular weight; $v_m$, $v_l$ and $v_t$ are the mean sound velocity, the longitudinal velocity, and the transverse sound velocity, respectively. The longitudinal and transverse sound velocities are obtained from Eq. (S4) after knowing the bulk ($B$) and shearing moduli ($G$) of the structure. For crystal structure, $B$ and $G$ can be calculated from the independent elastic constants. All calculations are performed using the procedure implemented in VASPkit program [2].

Thermal expansion coefficient (TEC) is given by Eq. (S5) [3].

$$\alpha = \frac{\Delta V}{\Delta T \cdot V} \quad (S5)$$

And the quasi-harmonic approximation (QHA) was used to calculate TEC [1]. The total Helmholtz free energy is given by Eq. (S6).

$$F(V,T) = E_0 + \frac{1}{2}\int_0^{\omega_{max}} g(\omega)\hbar\omega d\omega + kT\int_0^{\omega_{max}} g(\omega)\ln(1 - \exp(-\hbar\omega/kT))d\omega \quad (S6)$$

Where, $E_0$ is the total energy at 0 K, $\frac{1}{2}\int_0^{\omega_{max}} g(\omega)\hbar\omega d\omega$ is the zero point energy and $kT\int_0^{\omega_{max}} g(\omega)\ln(1 - \exp(-\hbar\omega/kT))d\omega$ is the Helmholtz free energy of lattice vibrations. For using Eq (S6) to evaluate the vibrational free energy, the phonon density of states (g($\omega$)) must be properly normalized to 3$N$, and $N$ represents the total number of atoms in the unit cell. By fitting EOS at different temperature, the equilibrium volume at different temperature can be obtained. The equilibrium volume at different temperature can be used to calculate the TEC curve.

Table SI PAW pseudopotentials employed in the DFT calculations for the relevant elements.

| | potential file for LDA | potential file for GGA and Meta-GGA |
|---|---|---|
| Ti | PAW Ti 03Oct2001 | PAW_PBE Ti 08Apr2002 |
| V | PAW V_sv 02Aug2007 | PAW_PBE V_sv 02Aug2007 |
| Cr | PAW Cr_pv 02Aug2007 | PAW_PBE Cr_pv 02Aug2007 |
| Fe | PAW Fe 03Mar1998 | PAW_PBE Fe 06Sep2000 |
| Co | PAW Co 26Mar2009 | PAW_PBE Co 02Aug2007 |
| Ni | PAW Ni 02Aug2007 | PAW_PBE Ni 02Aug2007 |
| Cu | PAW Cu_pv 19Apr2000 | PAW_PBE Cu 22Jun2005 |
| Y | PAW Y_sv 25May2007 | PAW_PBE Y_sv 25May2007 |
| Zr | PAW Zr_sv 04Jan2005 | PAW_PBE Zr_sv 04Jan2005 |
| Nb | PAW Nb_pv 09Jan2002 | PAW_PBE Nb_pv 08Apr2002 |
| Mo | PAW Mo_sv 29Jan2005 | PAW_PBE Mo_sv 02Feb2006 |
| Ru | PAW Ru_pv 28Jan2005 | PAW_PBE Ru_pv 28Jan2005 |
| Pd | PAW Pd 04Jan2005 | PAW_PBE Pd 04Jan2005 |
| Ag | PAW Ag 06Sep2000 | PAW_PBE Ag 02Apr2005 |
| Hf | PAW Hf_pv 17Apr2000 | PAW_PBE Hf_pv 06Sep2000 |
| Ta | PAW Ta_pv 07Sep2000 | PAW_PBE Ta_pv 07Sep2000 |
| W | PAW W 19Jan2001 | PAW_PBE W 08Apr2002 |
| Re | PAW Re 21Jan2003 | PAW_PBE Re 17Jan2003 |
| Os | PAW Os 21Jan2003 | PAW_PBE Os 17Jan2003 |
| Ir | PAW Ir 10Feb1998 | PAW_PBE Ir 06Sep2000 |
| Pt | PAW Pt_new 15Nov2007 | PAW_PBE Pt 04Feb2005 |
| Au | PAW Au 04Oct2007 | PAW_PBE Au 04Oct2007 |

Table SII Fitting parameters of Vinet EOS at 0 K corrected with zero point energy: Equilibrium total energy $E_0$ (eV), equilibrium volume $V_0$ (in Å$^3$), bulk modulus $B_0$ (in GPa) and its pressure derivative $B_1$ (dimensionless) by LSDA.

| Phase | $E_0$ (eV) | $V_0$ (Å$^3$) | $B_0$ (GPa) | $B_1$ |
|---|---|---|---|---|
| Ti | -17.05 | 31.87 | 133.01 | 3.58 |
| V  | -10.04 | 12.62 | 227.41 | 4.50 |
| Cr | -10.62 | 10.85 | 302.09 | 4.32 |
| Fe | -9.22  | 10.39 | 255.72 | 4.57 |
| Co | -16.24 | 19.95 | 268.28 | 1.68 |
| Ni | -6.59  | 9.87  | 264.43 | 4.51 |
| Cu | -4.74  | 10.95 | 183.00 | 5.44 |
| Y  | -13.89 | 59.64 | 43.85  | 3.09 |
| Zr | -18.58 | 43.67 | 103.31 | 3.13 |
| Nb | -11.15 | 17.39 | 194.86 | 3.94 |
| Mo | -12.13 | 15.14 | 289.62 | 4.20 |
| Ru | -21.38 | 26.26 | 365.82 | 4.81 |
| Pd | -6.64  | 14.17 | 229.60 | 5.15 |
| Ag | -3.78  | 16.06 | 136.87 | 5.58 |
| Hf | -21.42 | 41.65 | 121.55 | 3.41 |
| Ta | -12.83 | 17.30 | 219.71 | 3.78 |
| W  | -14.07 | 15.31 | 345.65 | 4.50 |
| Re | -27.66 | 28.67 | 414.16 | 4.53 |
| Os | -25.30 | 27.47 | 452.99 | 4.88 |
| Ir | -10.33 | 13.93 | 403.61 | 5.06 |
| Pt | -7.62  | 14.73 | 315.96 | 4.84 |
| Au | -4.44  | 16.65 | 191.30 | 5.45 |

Table SIII Fitting parameters of Vinet EOS at 0 K corrected with zero point energy: Equilibrium total energy $E_0$ (eV), equilibrium volume $V_0$ (in Å$^3$), bulk modulus $B_0$ (in GPa) and its pressure derivative $B_1$ (dimensionless) by the Perdew-Burke-Ernzerhof (PBE) GGA.

| Phase | $E_0$ (eV) | $V_0$ (Å$^3$) | $B_0$ (GPa) | $B_1$ |
|---|---|---|---|---|
| Ti | -15.51 | 34.26 | 116.63 | 3.70 |
| V | -9.09 | 13.44 | 193.80 | 6.24 |
| Cr | -9.50 | 11.53 | 258.25 | 4.34 |
| Fe | -8.23 | 11.40 | 185.23 | -0.96 |
| Co | -14.06 | 21.65 | 214.26 | 1.34 |
| Ni | -5.45 | 10.68 | 209.99 | 6.38 |
| Cu | -4.74 | 10.95 | 185.51 | 4.42 |
| Y | -12.86 | 65.59 | 40.99 | 3.04 |
| Zr | -17.03 | 46.83 | 93.61 | 3.31 |
| Nb | -10.09 | 18.33 | 173.38 | 4.07 |
| Mo | -10.93 | 15.84 | 259.42 | 4.10 |
| Ru | -18.46 | 27.54 | 311.95 | 4.92 |
| Pd | -5.36 | 15.26 | 172.84 | 5.53 |
| Ag | -2.72 | 17.84 | 91.03 | 5.91 |
| Hf | -19.84 | 44.82 | 113.93 | 3.67 |
| Ta | -11.81 | 18.30 | 197.43 | 3.44 |
| W | -12.86 | 15.93 | 309.45 | 4.35 |
| Re | -24.96 | 29.74 | 373.70 | 4.76 |
| Os | -22.42 | 28.56 | 397.59 | 4.91 |
| Ir | -8.85 | 14.52 | 348.59 | 5.20 |
| Pt | -6.20 | 15.51 | 263.00 | 5.64 |
| Au | -3.22 | 17.97 | 139.07 | 6.12 |

Table SIV Fitting parameters of Vinet EOS at 0 K corrected with zero point energy: Equilibrium total energy $E_0$ (eV), equilibrium volume $V_0$ (in Å³), bulk modulus $B_0$ (in GPa) and its pressure derivative $B_1$ (dimensionless) by the Perdew-Burke-Ernzerhof revised for solid (PBEsol) GGA.

| Phase | $E_0$ (eV) | $V_0$ (Å³) | $B_0$ (GPa) | $B_1$ |
|---|---|---|---|---|
| Ti | -16.56 | 33.09 | 125.10 | 3.71 |
| V  | -9.73  | 12.97 | 211.55 | 4.53 |
| Cr | -10.23 | 11.13 | 282.98 | 4.37 |
| Fe | -8.91  | 10.81 | 218.77 | 10.18 |
| Co | -15.46 | 20.69 | 246.77 | 1.54 |
| Ni | -6.13  | 10.21 | 240.81 | 4.89 |
| Cu | -4.33  | 11.36 | 168.65 | 3.35 |
| Y  | -13.63 | 62.51 | 42.26  | 2.33 |
| Zr | -18.15 | 45.01 | 98.86  | 3.19 |
| Nb | -10.82 | 17.73 | 186.56 | 3.44 |
| Mo | -11.77 | 15.37 | 279.05 | 4.05 |
| Ru | -20.33 | 26.65 | 347.77 | 4.87 |
| Pd | -6.10  | 14.52 | 207.39 | 5.48 |
| Ag | -3.31  | 16.66 | 118.31 | 5.56 |
| Hf | -21.00 | 43.11 | 120.10 | 3.82 |
| Ta | -12.55 | 17.69 | 208.76 | 3.86 |
| W  | -13.72 | 15.50 | 333.73 | 4.66 |
| Re | -26.86 | 29.00 | 401.82 | 4.44 |
| Os | -24.40 | 27.82 | 434.91 | 5.08 |
| Ir | -9.82  | 14.09 | 388.66 | 5.11 |
| Pt | -7.07  | 14.94 | 301.23 | 5.01 |
| Au | -3.92  | 17.02 | 174.37 | 5.65 |

Table SV Fitting parameters of Vinet EOS at 0 K corrected with zero point energy: Equilibrium total energy $E_0$ (eV), equilibrium volume $V_0$ (in Å$^3$), bulk modulus $B_0$ (in GPa) and its pressure derivative $B_1$ (dimensionless) by the revised-regularized strongly constrained and appropriately normed (r$^2$SCAN) metaGGA.

| Phase | $E_0$ (eV) | $V_0$ (Å$^3$) | $B_0$ (GPa) | $B_1$ |
|---|---|---|---|---|
| Ti | -25.72 | 34.05 | 123.55 | 3.63 |
| V | -14.13 | 13.32 | 193.77 | 3.88 |
| Cr | -14.80 | 11.36 | 275.78 | 4.21 |
| Fe | -14.35 | 11.77 | 201.40 | 2.17 |
| Co | | | | |
| Ni | -11.82 | 10.42 | 226.44 | 5.55 |
| Cu | -10.82 | 11.49 | 162.80 | 6.95 |
| Y | -40.53 | 66.99 | 40.90 | 3.13 |
| Zr | -45.49 | 47.02 | 95.39 | 3.25 |
| Nb | -24.72 | 18.34 | 173.72 | 4.55 |
| Mo | -26.07 | 15.71 | 274.99 | 4.18 |
| Ru | -50.85 | 27.12 | 337.50 | 4.82 |
| Pd | -22.94 | 14.98 | 184.55 | 5.62 |
| Ag | -21.36 | 17.32 | 104.37 | 5.97 |
| Hf | -90.30 | 44.24 | 117.45 | 3.98 |
| Ta | -48.58 | 18.10 | 205.69 | 4.20 |
| W | -51.47 | 15.72 | 331.20 | 4.48 |
| Re | -104.89 | 29.41 | 393.76 | 4.38 |
| Os | -106.01 | 28.10 | 434.73 | 4.94 |
| Ir | -52.37 | 14.30 | 377.39 | 5.17 |
| Pt | -51.50 | 15.33 | 272.28 | 5.48 |
| Au | -50.59 | 17.59 | 153.47 | 6.05 |

Table SVI For transition metals with hexagonal structures, the ratios of lattice constant c to a calculated by the four functionals.

| Phase | LDA | PBE | PBEsol | r$^2$SCAN |
|:-:|:-:|:-:|:-:|:-:|
| Ti | 1.58 | 1.58 | 1.58 | 1.58 |
| Co | 1.61 | 1.62 | 1.62 | |
| Y  | 1.56 | 1.55 | 1.55 | 1.55 |
| Zr | 1.62 | 1.60 | 1.61 | 1.59 |
| Ru | 1.58 | 1.58 | 1.58 | 1.58 |
| Hf | 1.58 | 1.58 | 1.58 | 1.58 |
| Re | 1.61 | 1.61 | 1.61 | 1.62 |
| Os | 1.58 | 1.58 | 1.58 | 1.58 |

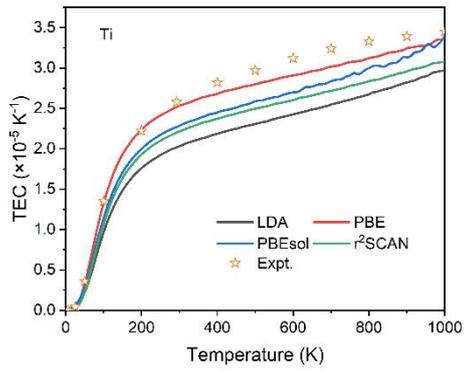

(a)

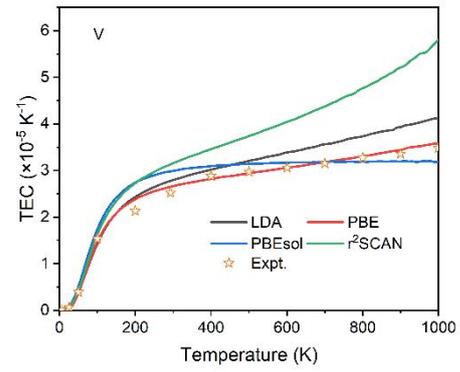

(b)

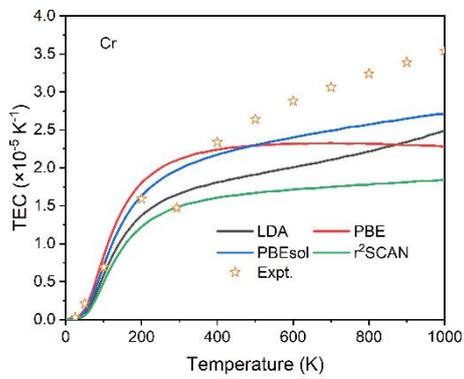

(c)

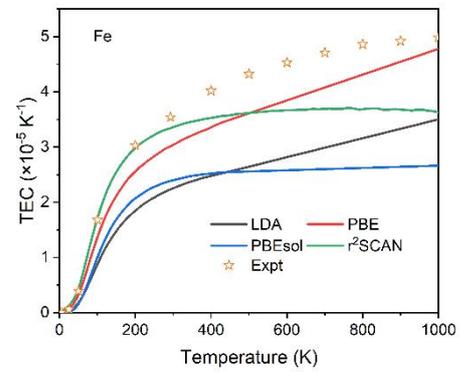

(d)

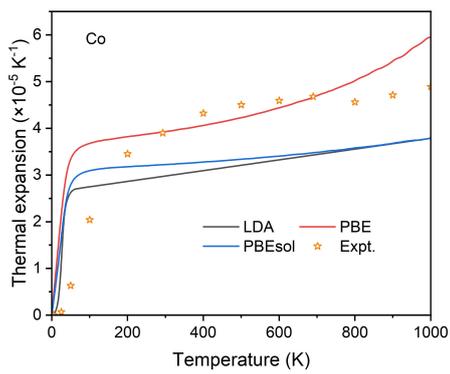

(e)

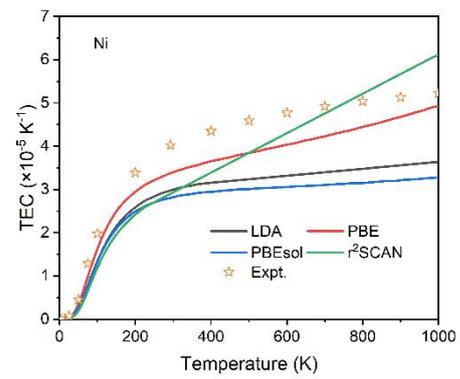

(f)

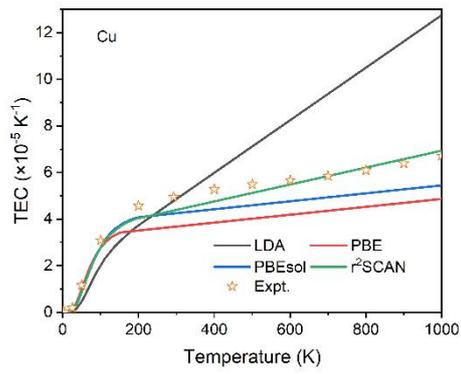

(g)

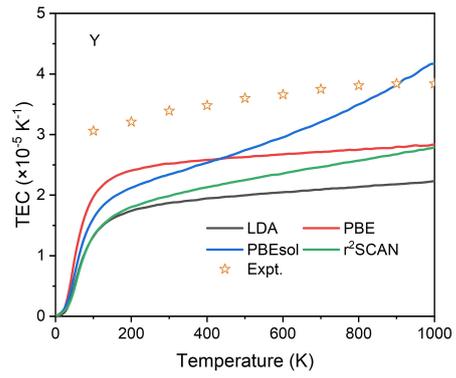

(h)

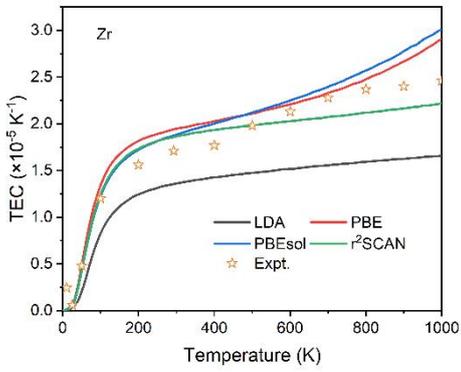

(i)

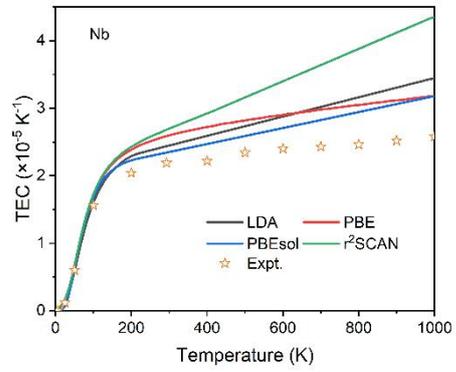

(j)

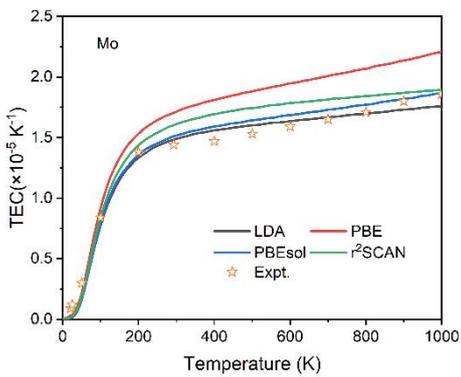

(k)

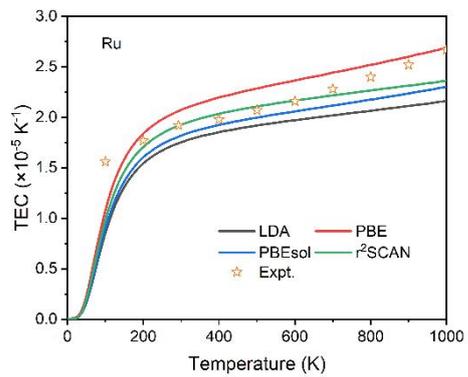

(l)

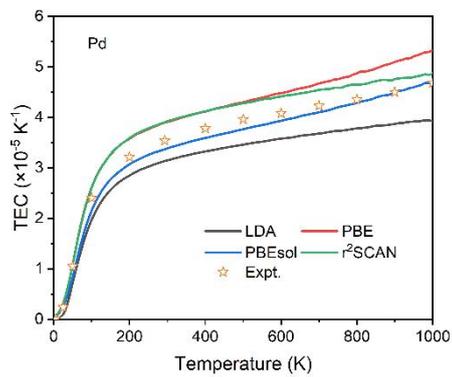

(m)

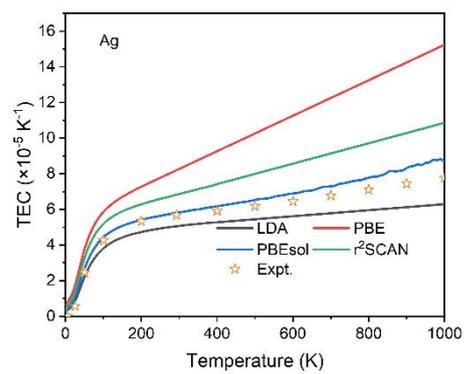

(n)

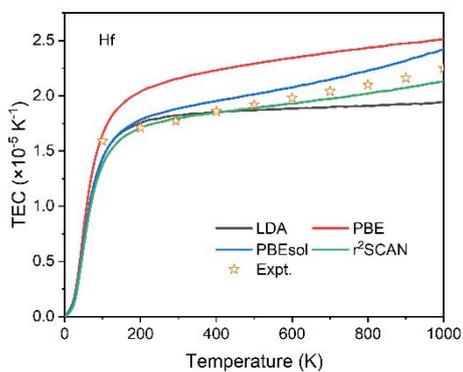

(o)

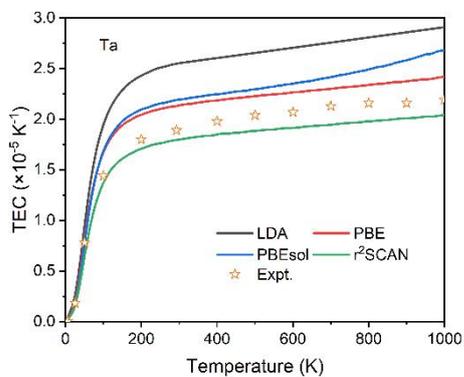

(p)

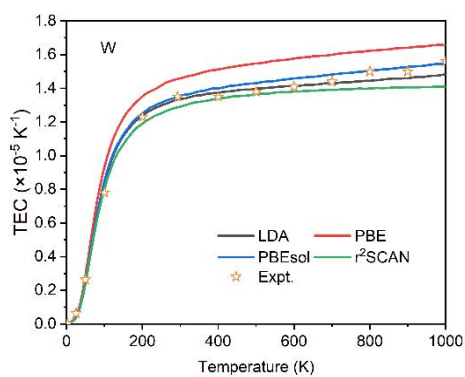

(q)

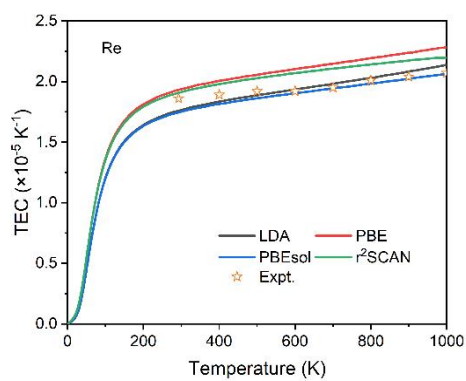

(r)

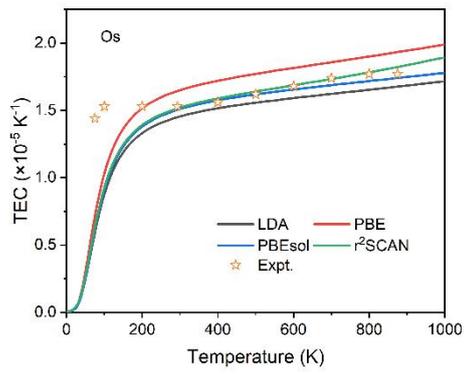

(s)

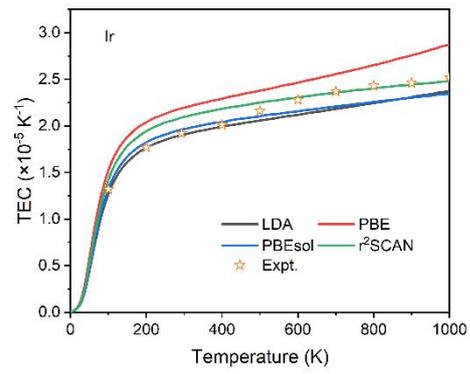

(t)

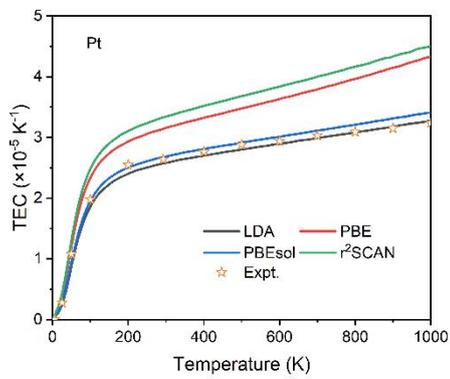

(u)

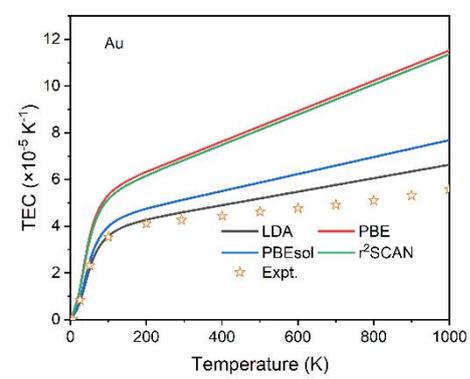

(v)

Figure S1 The calculated volumetric thermal expansion coefficient (TEC) of 3d, 4d and 5d transition metals using various semilocal exchange-correlation functionals, in comparisons with the experimental data [2].

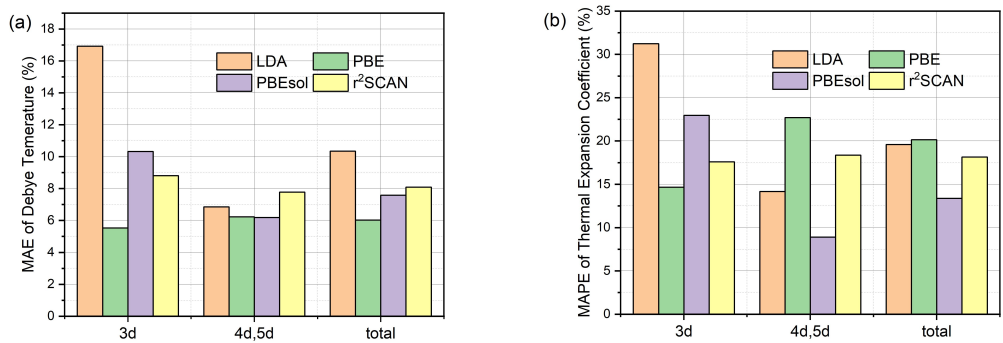

Figure S2 The mean percentage errors of Debye temperature (a) and volumetric TECs (b) of transition metals, obtained from semilocal functionals.

References

[1] Y. Ding and B. Xiao, 'Thermal expansion tensors, Grüneisen parameters and phonon velocities of bulk MT2 (M = W and Mo; T = S and Se) from first principles calculations', RSC Advances, 5, 18391 (2015). https://doi.org/10.1039/C4RA16966B

[2] V. Wang, N. Xu, J.C. Liu, G. Tang, W.T. Geng, 'VASPKIT: A User-Friendly Interface Facilitating High-Throughput Computing and Analysis Using VASP Code', Comput. Phys. Commun. 267, 108033 (2021). https://doi.org/10.1016/j.cpc.2021.108033

[3] Y. S. Touloukian, R. K. Kirby, R. E. Taylor, et al. 'Thermal Expansion: Metallic Elements and Alloys', Thermophysical Properties of Matter, 12(1975).